\documentclass{emulateapj}
\usepackage{apjfonts}

\newcommand\chandra{{\it Chandra}}
\newcommand\eg{{e.g.}}
\newcommand\etal{{et al.}}
\newcommand\msun{{\rm M_\odot}}
\newcommand\ergps{{\rm erg\ s^{-1}}}

\begin{document}

\title{The cluster-scale AGN outburst in Hydra A}

\author{P. E. J. Nulsen\altaffilmark{1,2},
  B. R. McNamara\altaffilmark{3}, M. W. Wise\altaffilmark{4},
  L. P. David\altaffilmark{1}}
\altaffiltext{1}{Harvard-Smithsonian Center for Astrophysics, 60
  Garden Street, Cambridge, MA 02138} 
\altaffiltext{2}{On leave from the University of Wollongong}
\altaffiltext{3}{Department of Physics and Astronomy, Ohio University,
  Clippinger Laboratories, Athens, OH 45701}
\altaffiltext{4}{Center for Space Research, Building NE80-6015,
  Massachusetts Institute of Technology, Cambridge, MA 02139}

\begin{abstract}

Deep \chandra\ observations of the Hydra A Cluster reveal a feature in
the X-ray surface brightness that surrounds the 330 MHz radio lobes of
the AGN at the cluster center.  Surface brightness profiles of this
feature and its close association with the radio lobes argue strongly
that it is a shock front driven by the expanding radio lobes.  The
\chandra\ image also reveals other new structure on smaller scales
that is associated with the radio source, including a large cavity and
filament.  The shock front extends 200 -- 300 kpc from the AGN at the
cluster center and its strength varies along the front, with Mach
numbers in the range $\sim 1.2$ -- 1.4.  It is stronger where it is
more distant from the cluster center, as expected for a shock driven
by expanding radio lobes.  Simple modeling gives an age for the shock
front $\sim 1.4\times10^8$ y and a total energy driving it of $\sim
10^{61}$ erg.  The mean mechanical power driving the shock is
comparable to quasar luminosities, well in excess of that needed to
regulate the cooling core in Hydra A.  This suggests that the feedback
regulating cooling cores is inefficient, in that the bulk of the
energy is deposited beyond the cooling core.  In that case, a
significant part of cluster ``preheating'' is a byproduct of the
regulation of cooling cores.

\end{abstract}


\section{Introduction}

The X-ray emitting gas at the centers of many cooling flow clusters is
cooler than the surrounding gas and has cooling times in the range
$10^8$ -- $10^9$ y.  Despite this, little of the gas cools below $\sim
1$ keV \citep[\eg][]{ppk01,tkp01,pkp03,ktp04}.  The main issue for
cooling flows has become to determine what heat source makes up for
radiative losses from the gas.  Many possibilities have been proposed,
including thermal conduction \citep[\eg][]{nm01}, energy released by
mergers \citep[\eg][]{mbl04} and heating by an active galactic nucleus
\citep[AGN;][]{tb93,td97}.  However, maintaining gas for many cooling
times, while preventing it from cooling to low temperatures or being
heated until its cooling time is comparable to its age, requires a
fine balance between heating and cooling.  Since cooling flows are
common \citep{f94}, they cannot be a transient phenomenon, and
plausible mechanisms for maintaining the gas with short cooling times
must almost certainly involve feedback.  While they face some
difficulties \citep[\eg][]{fmn01}, AGN heating models provide a
natural feedback mechanism, making them the strongest candidate to
solve this cooling flow problem.

There is clear evidence for some AGN heating in cooling flows.  Large
galaxies are prevalent at cluster centers, and \citet{b90} has shown
that cD galaxies at the centers of cooling flows have a high incidence
of radio activity.  High resolution X-ray images reveal that radio
lobes fed by AGN have created cavities in the hot gas in a growing
number of clusters \citep{bvf93,cph94,mwn00,fse00,mwn01,bsm01,scd01,%
hcr02,ywm02} and also in the hot interstellar medium of individual
elliptical galaxies \citep{fj01,kfc04}.  \citet{csf02} pointed out
that the enthalpy of a cavity (bubble) can be thermalized as it rises
buoyantly, providing a heat source for the gas.  Simple arguments show
that the enthalpy is thermalized in the wake of a rising bubble,
provided that the bubble is not too large compared to its distance
from the cluster center \citep{brm04}.  Using optimistic estimates for
the heating rate, \citet{brm04} analyzed data for all cavities they
could identify in the \chandra\ archive and found that, while the
heating rate due to bubbles can be significant, it typically falls
several times short of radiative losses.

Standard models of jet-fed radio lobes \citep[\eg][]{s74,hrb98} posit
the existence of a ``cocoon shock'' that surrounds the rapidly
expanding radio lobes.  Although there is commonly a bright rim around
the radio lobe cavities in X-ray clusters, this is composed of cool
(low entropy) gas, close to local pressure equilibrium
\citep[\eg][]{ndm02}, so that most cavities are not strongly over
pressured \citep[Cen A is a noteworthy exception,][]{kvf03}.  However,
evidence has begun to emerge for weak shocks generated by the radio
lobes in cooling flows.  \citet{fsa03} found ripples in the surface
brightness of the Perseus cluster, which they interpret as weak shocks
(sound waves) generated as the expansion rate of the lobes varies.
They argue that heating due to viscous dissipation of sound can make
up for radiative losses from the Perseus cooling flow.  This process
has been simulated by \citet{rbb04}.  \citet{ywm02} found a weak shock
surrounding M87 in the Virgo Cluster, and \citet{fnh04} found evidence
of further shocks in a deeper X-ray image.  They determined the energy
of the outburst that drove the best defined shock, using the result to
show that heating due to weak shocks is sufficient to stop the cooling
flow in M87.

Here we report the discovery of a weak shock front generated by an AGN
outburst in the Hydra A Cluster.  Hydra A has well known radio lobe
cavities \citep{mwn00,dnm01,ndm02}, which surround the 1.4 GHz inner
radio lobes and extend to $\sim40''$ (40 kpc) from the cluster center.
By contrast, the weak shock front discussed here extends up to
$\sim6'$ from the cluster center, surrounding the much larger low
frequency radio lobes \citep{lct04} and representing a much more
energetic phenomenon.

In section \ref{sec:obs} we give details of the observations and data
reduction.  In section \ref{sec:struct} we discuss the main features
of the new deep image of Hydra A and in section \ref{sec:shock} we
discuss the evidence for the shock front and estimate the energy of
the outburst that drove it.  The implications of this shock for
cooling flows and clusters in general are discussed in section
\ref{sec:disc}.  We assume flat $\Lambda$CDM, with $H_0 = 70\rm\ km\
s^{-1}\ Mpc^{-1}$ and $\Omega_{\rm m} = 0.3$, giving a scale of
$1.05\rm\ kpc\ arcsec^{-1}$ at the redshift of Hydra A, $z = 0.0538$.

\section{Observations and data reduction} \label{sec:obs}

Hydra A was observed with \chandra\ for 98.2 ksec on 13 Jan 2004, with
ACIS-S at the aim point in VFAINT mode (OBSID 4969).  A z-sim offset
of 5 mm was used to keep the cluster center close to the optical axis,
while extending the field of view to the north to cover the northern
low frequency radio lobe \citep{lct04}.  The event list was screened
to removed ASCA grades 1, 5 and 7, and bad pixels in the standard
manner.  A large background flare made it necessary to set the mean
background count rate manually when using lc\_clean to remove periods
of high particle background, but the mean count rate in ACIS S1 after
cleaning is close to the expected
value\footnote{http://cxc.harvard.edu/contrib/maxim/acisbg/data/README}.
After cleaning, 66.4 ksec of good data were left.  The data were
processed to correct for time dependence of the ACIS
gain\footnote{http://hea-www.harvard.edu/{${\sim}$}alexey/acis/tgain/}.
Data were filtered according to the prescription of
Vikhlinin\footnote{http://cxc.harvard.edu/cal/Acis/Cal\_prods/vfbkgrnd/}
to reduce particle background.  Background event files were created by
processing the standard ACIS background files in the same manner as
the data.  Point sources were identified manually for removal from
spectra and surface brightness profiles.  When extracting spectra,
ARF's and RMF's for extended regions are weighted by the number of
events in each subregion.  ARF's are corrected to allow for the
reduction in low energy response due to the build up of contaminant on
the ACIS filters.

\section{X-ray structure of Hydra A} \label{sec:struct}

\begin{figure}
\includegraphics[width=\linewidth]{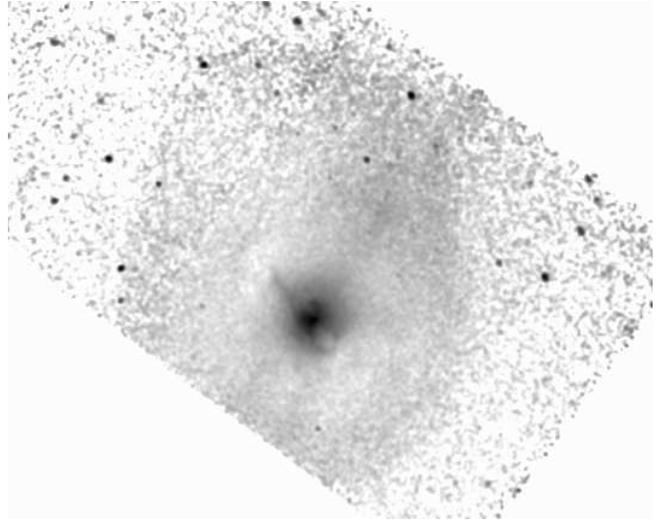}
\caption{\chandra\ 0.5 -- 7.5 keV image of Hydra A.  The image from
  the S2 and S3 chips has been background-subtracted,
  exposure-corrected, smoothed with a $2''$ gaussian and divided by a
  spherical beta model centered on the AGN.  The chips are $8'\!.4$
  wide.  The cavities of \citet{mwn00} can be seen close to the bright
  AGN.  The cavity $1'$ -- $2'$ northeast of the AGN (cavity C), at
  the end of the bright filament, occurs where the 330 MHz radio jet
  bends sharply to the northwest (see Fig.~\ref{fig:shrad}).  The
  shock front surrounding the radio lobes of Fig.~\ref{fig:shrad} is
  visible, particularly to the east and north.}\label{fig:smff}
\end{figure}

\begin{figure}
\includegraphics[width=\linewidth]{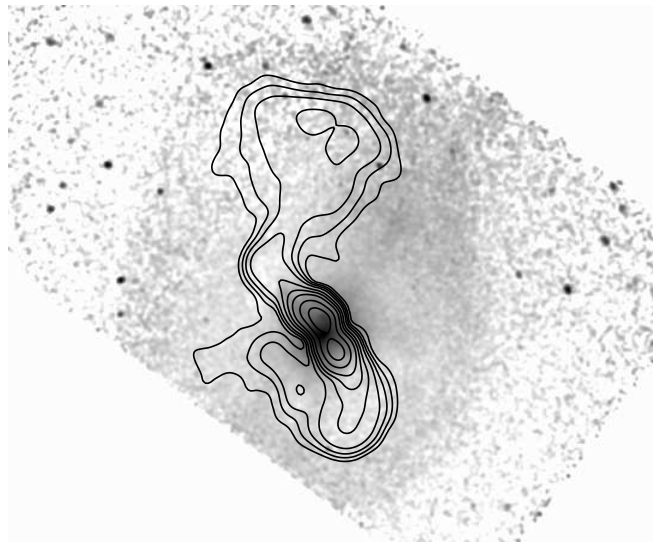}
\caption{X-ray image of Hydra A with 330 MHz radio contours.  The
  image of Fig.~\ref{fig:smff} is shown with logarithmically spaced
  contours from the 330 MHz radio map.  The two closed contours near
  the center of the map are at the position of the 1.4 GHz radio
  lobes, marking the locations of the inner X-ray cavities that can be
  seen in Fig.~\ref{fig:smff}.  }\label{fig:shrad}
\end{figure}

Fig.~\ref{fig:smff} shows a background-subtracted, exposure-corrected
and smoothed image of Hydra A in the 0.5 -- 7.5 keV band.  In order to
make structures visible over a wide range of radius, the image has
been divided by a spherical beta model, with a core radius of $98''$,
$\beta=0.6$ and centered on the AGN.  The field is covered by ACIS S3
and part of S2 (to the northeast).  For scale, the chips are $8'\!.4$
wide.  The southwest cavity of \citet{mwn00} can be seen $\sim0'\!.5$
to the southwest of the bright AGN.  The corresponding radio cavity to
the northeast is also visible, though less clearly defined.  A $\sim
1'$ bright filament stretches from the inner cavity to about the
center of a larger outer cavity in the northeast (cavity C).  The
association between the $1'$ bright filament to the northeast and the
radio jet of Hydra A is reminiscent of the southwest filament in M87
\citep{fnh04}, suggesting a similar origin.  The filament in Hydra A
is roughly twice as long as that in M87, but the radio source in Hydra
A is a lot larger too.  Cavity C has a radius $\sim0'\!.6$ (40 kpc)
and occurs at the sharp bend in the radio jet leading to the northern
330 MHz radio lobe (see Fig.~\ref{fig:shrad}).  There is also a
deficit in the X-ray emission $\sim 2'$ to the southwest of the AGN,
associated with the southern 330 MHz radio lobe.

There is an edge in X-ray surface brightness, running from
$\sim4'\!.3$ east to $\sim6'$ north of the AGN, that continues around
the AGN.  The northern part of this edge was noted previously by
\citet{fdj00} and by \citet{mvf02}.  Although it is less distinct to
the west, the feature can be traced all the way around the AGN, except
where it falls off the detectors in the south.  We will interpret the
feature as a shock front and refer to it this way from here on.
Fig.~\ref{fig:shrad} shows the same X-ray image as Fig.~\ref{fig:smff}
together with contours of the 330 MHz radio map of \citet{lct04}.  The
northern radio lobe lies close inside the northern part of the shock
front and has a similar shape.  Although the X-ray image of the shock
front is truncated to the south, enough of it is visible to see that,
along with the 330 MHz radio lobe, it lies closer to the AGN in the
south.  The correspondence between the shapes of the radio lobes and
the shock front supports the interpretation of this surface brightness
feature as a shock front driven by the outburst that formed the radio
lobes.

\section{Shock models for Hydra A} \label{sec:shock}

\subsection{Shock brightness profiles}

\begin{figure}
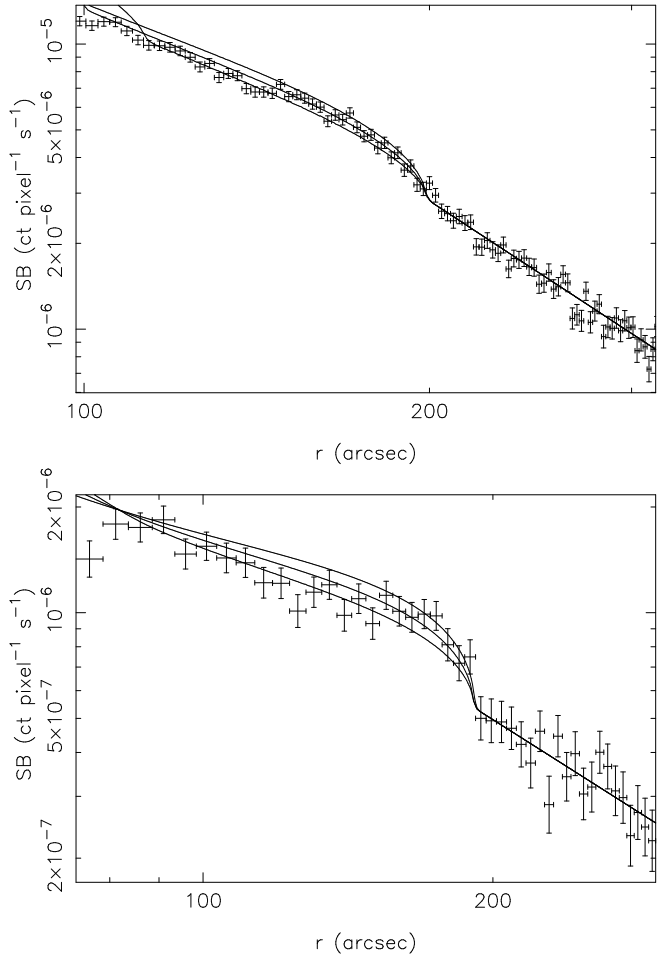

\centerline{\includegraphics[height=\linewidth,angle=270]{pubsbprof.ps}}
\centerline{\includegraphics[height=\linewidth,angle=270]{pubsbprofne.ps}}
\caption{Surface brightness profiles of the shock front in Hydra A.
  {\it Upper panel:} The radial surface brightness profile measured in
  the sector from PA $240^\circ$ to $300^\circ$, to the west of the
  AGN, and in the energy range 0.6 -- 7.5 keV.  Surface brightness
  errors are 1 sigma statistical errors.  Radial error bars show the
  limits of the bins.  The smooth curves show surface brightness
  profiles for shock models with Mach numbers of 1.15, 1.19 and 1.23,
  from bottom to top.  Models are scaled to match the observed surface
  brightness outside the shock.  {\it Lower panel:} 0.6 -- 7.5 keV
  surface brightness profile of the shock front in the northeast,
  measured in a sector from PA $20^\circ$ to $50^\circ$ and centered
  $2'\!.9$ north of the AGN.  Models are shown for shocks with Mach
  numbers of 1.26, 1.34 and 1.42.}
  \label{fig:sbw}
\end{figure}

The upper panel of Fig.~\ref{fig:sbw} shows the surface brightness
profile of the region to the west of Hydra A, in a sector centered on
the AGN, from PA $240^\circ$ to $300^\circ$.  Point sources were
eliminated, background subtracted and the resulting profile exposure
corrected.  The abrupt change in slope at a radius of $202''$ (211
kpc) is due to the feature that we identify as a shock.  Outside this,
the surface brightness is well fitted by the power law, $r^{-\alpha}$
with $\alpha = 2.65\pm0.17$ (90\% confidence).  A surface brightness
profile was also extracted in the region to the northeast of the AGN,
where the edge in the surface brightness is most prominent.  This is
shown in the lower panel of Fig.~\ref{fig:sbw}.  In order to match the
curvature of the radial bins to that of the front, the sector was
centered at a point approximately $2'\!.9$ north of the AGN (at J2000
$\rm RA, \delta = 9^h 18^m 07^s\!\!.30,\, -12^\circ 02' 48''\!\!.4$),
with the range of PA $20^\circ$ to $70^\circ$.  Although the data are
noisy, the edge is evident in the surface brightness profile at a
radius of about $200''$ (measured from the center of the sector).

The break in surface brightness in the northeast is greater than that
in the west, showing that the shock is stronger in the northeast.
This is consistent with expectations for radio lobe driven shocks.
Shock strength is determined by the ratio of postshock to preshock
pressure.  The preshock pressure is significantly lower in the north,
where the shock is farthest from the cluster center.  To a first
approximation, the pressure rise behind the shock is uniform around
the front (assuming that jet momentum is insignificant on such large
scales).  Thus, the pressure jump will be greater at points on the
shock front that are further from the cluster center.

The shock front lies $\sim3'\!.4$ west of the AGN and $\sim4'\!.3$ to
the east, implying a significant east-west asymmetry in shock speed.
This is most likely due to asymmetry in the medium carrying the shock
(as opposed to the outburst driving it) and probably the result of
higher gas density to the west.  If the cluster were spherically
symmetric, the shock front would be running through denser gas in the
west than in the east.  Any pre-existing density asymmetry that slowed
the propagation of the shock to the west would augment this density
difference.  If postshock pressure is much the same in the east and
west, then the shock will be weaker in the west, where it propagates
through higher pressure gas.  The western shock also appears
superimposed on a higher ``background'' of emission from undisturbed
cluster gas.  Together, these effects make the shock less distinct in
the X-ray image to the west than to the east, but it is clearly
visible in the upper panel of Fig.~\ref{fig:sbw}.

\subsection{Hydrodynamic models}

To quantify the shock, we have used a spherically symmetric,
hydrodynamic model of a point explosion at the center of an initially
isothermal, hydrostatic atmosphere.  For the shock front in the west,
the initial gas density profile is assumed to be a power law, $\rho(r)
\propto r^{-\eta}$, with $\eta = 1.82$, which makes the surface
brightness profile of the undisturbed gas consistent with the observed
surface brightness profile beyond the shock (see above).  The
gravitational field ($g \propto 1/r$) and gas temperature are
scaled to make the undisturbed atmosphere hydrostatic.  Surface
brightness profiles are determined from the model, assuming that the
temperature of the unshocked gas is 4 keV.  The {\it Chandra} 0.6 --
7.5 keV response was computed using XSPEC, with detector response
files appropriate for these observations and an absorbed mekal model.
The foreground column density was set to $4.94\times10^{20}\rm\
cm^{-2}$, the redshift to 0.0538 and the abundance to 0.3 times solar,
as appropriate for Hydra A (results are insensitive to these
parameters, including the preshock temperature).  The shock weakens as
the hydrodynamic model evolves and, since the initial conditions are
self-similar, the flow can be scaled radially to place the model shock
at the location of the observed shock.  Surface brightness is
scaled to match the observed profile in the unshocked region.

The three lines in the upper panel of Fig.~\ref{fig:sbw} show model
surface brightness profiles for shocks with Mach numbers of 1.15, 1.19
and 1.23 (increasing upward).  A Mach 1.2 shock gives a reasonably
good fit to the observed shock profile to the west.  Models were also
constructed for the shock to the northwest.  Here the preshock density
profile is flatter, requiring an initial density power law $\eta =
1.37$ (largely because the shock does not propagate radially in the
cluster).  Model surface brightness profiles are shown in the lower
panel of Fig.~\ref{fig:sbw} for Mach numbers of 1.26, 1.34 and 1.42.
The Mach number of the shock to the northeast $\sim1.3$ -- 1.4, again
confirming that the shock is stronger in the northeast than in the
west.

Our simple hydrodynamic model is not accurate well behind the shock.
The initial density profile is only well approximated as a power law
locally.  The shock front is clearly aspherical.  The outer radio
lobes lie close behind the shock front, so that they still have an
influence in driving the shock (if not, they would have been left
behind the rapidly moving front), violating our assumption that the
shock front is driven by a point explosion.  However, since the lobes
are not close to the shock front everywhere, it would also be
necessary to drop our assumption of spherical symmetry to make
significantly more accurate models.  Because of this, the added
complexity of models with energy fed continuously into the center of
the flow is not warranted.  Despite these shortcomings, the models do
provide reasonable fits to the surface brightness in the region of the
shock front.  In particular, they give reasonably accurate measures of
the Mach number (the main source of uncertainty is the assumption,
implicit in the spherical model, that we know the curvature of the
front along the line of sight).

\subsection{Physical parameters of the shock}

We base our estimates of outburst age and energy on the model for the
western shock, since the assumptions of spherical symmetry and a point
explosion are more appropriate in this region.  The age and energy of
the model outburst depend on the temperature and density of the
unshocked gas, which we determine by deprojection.  The sector to the
west of the AGN, in which the surface brightness profile of the upper
panel of Fig. \ref{fig:sbw} was measured, was divided into annular
regions having at least 4000 source counts in the 0.6 -- 7.5 keV band.
One boundary was fixed at the radius of the shock and others fitted
around this in order to avoid smoothing across the shock.  Spectra
extracted from these regions were used for the deprojection.  The
method is essentially as described in \citet{dnm01}.  The spectral
model for each ring (annular region) includes components for all outer
shells as well as the current one.  Only model parameters for the
current shell are varied when fitting a ring, so that parameters for
each spherical shell are determined in turn, working inward.  The
model component for each outer shell is weighted by the ratio of the
volumes of the intersections between that shell, and the current and
outer rings.  However, weights for the outermost shell were determined
on the assumption that the gas density follows the same power law as
the undisturbed gas in the shock model ($\rho \propto r^{-1.82}$),
from the inside edge of that shell to infinity, so allowing for the
remainder of the cluster.  Gas density is determined from the XSPEC
norm of each model component, assuming that the gas density is
constant in the corresponding shell.  For the outermost shell, the gas
density is determined at the center of the shell, under the
assumptions used to determine the weights for that shell.

Even with 4000 counts per spectrum, the errors for the deprojected
temperatures ($\sim 1$ keV) are too large to reveal the temperature
jump of the shock.  For a shock Mach number of 1.2, the temperature
jumps by $\sim20\%$ at the shock, but it falls rapidly behind the shock
due to adiabatic expansion.  As a result, the temperature rise in the
shell immediately inside the shock is only expected to be $\sim10\%$.
The preshock temperature determined from the deprojection is
approximately 4 keV and we use this value for further calculations.

\begin{figure}
\includegraphics[height=\linewidth,angle=270]{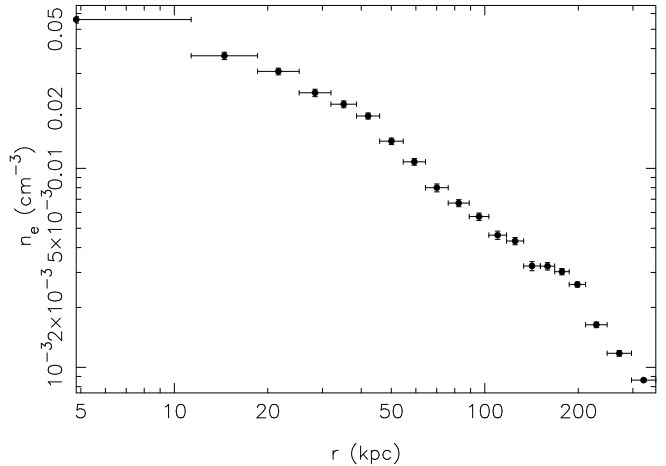}
\caption{Western electron density profile of Hydra A.  Deprojected
  electron density versus radius in the sector from PA $240^\circ$ to
  $300^\circ$.  The shock jump is at 211 kpc.  Density error bars are
  90\% confidence ranges, but some are hidden by the data point.}
  \label{fig:dep} 
\end{figure}

The deprojected electron density profile is shown in
Fig.~\ref{fig:dep}.  Electron densities are determined much more
accurately than temperatures, chiefly because the overall \chandra\
count rate is insensitive to gas temperature in the range of interest.
The density jump due to the shock at a radius of 211 kpc can be seen
clearly.  It is consistent with the $\sim30\%$ jump expected for a
Mach 1.2 shock.

Using the deprojected temperature and density of the unshocked gas to
normalize the physical parameters of the shock model, we find that the
age of the outburst is $t_{\rm s} = 1.4\times 10^8$ y and its total
energy is $E_{\rm s} = 9\times10^{60}$ erg.  The main source of
uncertainty in the age of the shock is the preshock temperature, since
the Mach number is not sensitive to the model (if the shocked cocoon
is axially symmetric, the size we measure is not affected by
projection).  Shock energy is proportional to the preshock
temperature.  It also depends on the density profile of the unshocked
gas.  There is a substantial geometric uncertainty, since the shock
front is clearly aspherical.  Our estimate uses the piece of the shock
that is nearest to the cluster center, underestimating the total
shocked volume, so tending to underestimate the total energy.  Lastly,
our estimate is affected by the inaccurate assumption of a point
explosion.  Despite all this, we expect the true energy of the
outburst to be no more than a factor $\sim 2$ astray.  The mean
mechanical power of the outburst is $P_{\rm s} = E_{\rm s} / t_{\rm s}
\simeq 2\times 10^{45}\ \ergps$.

\subsection{Shock vs Cold Front}

\begin{figure}
\includegraphics[height=\linewidth,angle=270]{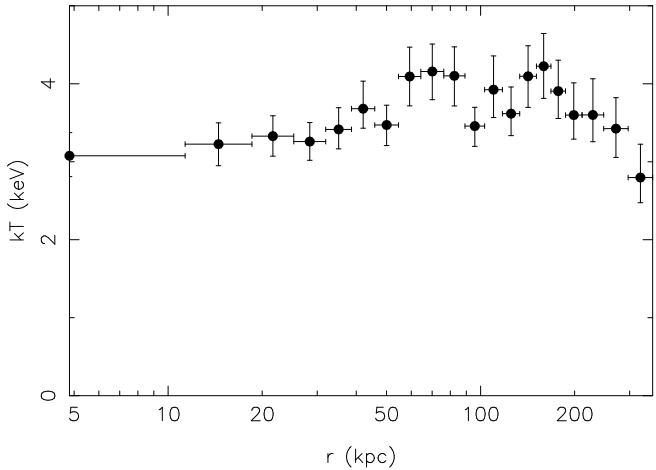}
\caption{Western temperature profile of Hydra A.  Projected
  temperature versus radius in the section from PA $240^\circ$ to
  $300^\circ$.  Error bars shows 90\% confidence ranges for the
  temperature.} \label{fig:tproj}
\end{figure}

Interpreting the shock feature as a cold front instead would be
difficult, although temperatures are not determined accurately enough
from the current data to categorically rule this out.  The density
discontinuity (Fig.~\ref{fig:dep}) shows that there is a front.  If it
is a cold front, the pressure must be continuous there.  Otherwise the
only physically reasonable interpretation is a shock.  For a density
decrease $\sim30\%$, the temperature must increase by $\sim30\%$ to
keep the pressure continuous.  Fig.~\ref{fig:tproj} shows projected
temperature versus radius for the sector where the deprojection was
done.  Temperatures were obtained by fitting absorbed mekal models,
with the absorbing column fixed at the foreground value ($N_{\rm H} =
4.94 \times 10^{20} \rm\ cm^{-2}$) and the abundance free.  If
anything, temperature decreases with radius at the front ($r = 211$
kpc).  The same is true at other PA's, where greater density steps
would require greater temperature steps.  According to the shock
model, the maximum rise in projected temperature should be $\sim5\%$,
confined to the two rings behind the shock.  This is consistent with
Fig.~\ref{fig:tproj}, although there is a mild negative temperature
gradient in the vicinity of the shock that is not included in our
simple shock model.  The association between the front and the radio
lobes is much better explained by a shock driven by an outburst than
by a cold front.  Furthermore, cold fronts are thought to be contact
discontinuities formed when a body of gas moves through gas of higher
entropy.  In that case, it would be difficult to form a closed cold
front, whereas a shock driven by an outburst at the cluster center is
expected to surround the AGN.  Altogether, the evidence strongly
favors interpretation of this feature as a shock front.

\section{Discussion} \label{sec:disc}

\subsection{Jet History}

X-ray measurements of shock properties provide a new tool for the
study of AGN outbursts and radio sources.  As discussed above, the
uncertainty in the age of the outburst, $\sim15\%$, is mainly due to
uncertainty in the the preshock temperature.  Shock strength is
determined by the overpressure of the radio lobes.  If it remains
constant, then, under the assumptions of our simple shock model, the
total energy of the shocked region scales as $E_{\rm s} \propto r_{\rm
s}^{3 - \eta} \propto t_{\rm s}^{3 - \eta}$, where $r_{\rm s}$ is the
shock radius and $t_{\rm s}$ is the time since the outburst.  Jet
power needs to vary with time as $t^{2 - \eta}$ to keep up with this.
The shock weakens or strengthens with time, depending on whether jet
power lags or leads this time dependence.  The fraction of the total
energy going into cavity enthalpy also depends on the overpressure of
the cavities.  Detailed modeling will constrain the history of the AGN
outburst.

Although there are two X-ray cavities visible along the northeastern
radio jet, none is evident at the location of the northern radio lobe.
However, the surface brightness is low there and has structure on the
scale of the lobe (including the shock front) that helps to mask any
cavity if it is present.  As found for the inner cavities
\citep{ndm02}, cavity C in the northeast is made easier to recognize
by the bright rim along its outer edge (Fig.~\ref{fig:smff}).  Without
this, the contrast of cavity C would only be $\sim 10\%$, making it
considerably harder to see.

The structure of the northeastern jet and the shock front suggest a
complex history for the radio source.  If the bulk of the northeastern
jet was filled with a connected body of radio plasma at the onset of
the current outburst, a compression wave (shock) would have been
launched simultaneously from its whole length.  If, as it is now, the
shock was strongest in the regions of lowest preshock pressure, near
to the outer lobes, it would have propagated farthest in those regions
\citep[as in the simulations of][]{rbb04}.  This is inconsistent with
the bulges in the shock front to the east and west of the center,
suggesting that this did not happen.  At the other extreme, the
outburst could have launched a new jet that has penetrated $\sim300$
kpc from the AGN to the northern radio lobe in $\sim1.4\times10^8$ y,
with a mean speed of $v_{\rm j} \simeq 2100\rm\ km\ s^{-1}$.  It
appears unlikely that jet momentum would have been significant for the
propagation of the jet beyond the location of cavity C, where it turns
through $\sim90^\circ$.  If so, the mean speed of the lobe is
excessive for a bubble that was driven most of the way by buoyancy
\citep[\eg][]{cbk01}.  The intermediate case, of a newly launched jet
connecting after a time into aging trails of radio plasma left by
earlier outbursts can avoid these two extremes.  It also requires more
modest asymmetries in gas properties than the pure jet model to
account for the north-south asymmetry of the radio structure and shock
front.

\subsection{Cooling Flow Feedback and Preheating}

It needs to be asked whether the outburst in Hydra A has any
connection to the cooling flow.  If so, then the gas that fuels it
must be cooled or cooling intracluster medium (ICM).  The line
emitting gas at the center of Hydra A might have been deposited by a
merger \citep{m95}, which could also be the source of fuel for the
outburst.  As discussed in the introduction, the strongest argument
for a connection between the outburst and the cooling flow is the
difficulty of maintaining cool cores without feedback, but this does
not rule out occasional mergers leading to larger than usual
outbursts.  On the other hand, large outbursts appear to play a
significant role in balancing radiative losses (see below).  In that
case, it would be surprising if they were not part of the feedback
process.  The correlation between cooling luminosity and cavity
heating power found by \citet{brm04} also suggests strongly that the
two are connected by feedback across the full range of systems.  The
outburst in Hydra A is three orders of magnitude more energetic than
that which produced the 14 kpc shock in M87 \citep[$\sim10^{58}$
erg,][]{fnh04}.  The outbursts responsible for the ripples in Perseus
\citep{fsa03} lie somewhere between these in energy.  MS0735.6+7421
\citep{mnw04} and Hercules A (Nulsen \etal, in preparation) both have
more energetic outbursts than Hydra A.  At the least, outbursts span a
large range of energies.  A large sample is required to determine
whether the large outbursts form a distinct class from those required
to regulate cooling cores.

In the absence of heat sources, radiating gas cools to low temperature
in approximately one cooling time, even after allowing for flow
induced by the cooling.  Thus, if the mass of gas inside radius $r$ is
$M_{\rm g}(r)$ and the cooling time of gas at radius $r$ is $t_{\rm
c}(r)$, a good approximation to the cooling rate when the gas between
$r_1$ and $r_2$ is cooling to low temperatures is \citep{fn79}
\begin{displaymath}
\dot M \simeq  {M_{\rm g}(r_2) - M_{\rm g}(r_1) \over t_{\rm c}(r_2) -
  t_{\rm c}(r_1)}.
\end{displaymath}
Applying this to the results of \citet{dnm01} for Hydra A shows that
the cooling rate would reach $100\ \msun\rm\ y^{-1}$ almost
immediately at the onset of cooling, and climb steadily thereafter.
Such high cooling rates have yet to demonstrated convincingly for any
cluster \citep[\eg][]{pkp03} and so must be rare.  This requires that,
if there is any period when the gas cools to low temperatures, it must
be brief compared to observed central cooling times.  Thus another
outburst must occur before, or very soon after, the central gas begins
to cool in $\sim4\times10^8$ y \citep{dnm01}.  The maximum time
between the last outburst and the next is therefore $t_{\rm max} \sim
5\times10^8$ y, and the minimum average power required for such
outbursts to prevent the gas from cooling is $P_{\rm min} \simeq
E_{\rm s} / t_{\rm min} \simeq 6\times10^{44}\ \ergps$, significantly
more than the $2.5\times10^{44}\ \ergps$ radiated from within 150 kpc
of the cluster center \citep{dnm01}.  The outburst in Hydra A cannot
be a very efficient process for regulating the cool core.  The shock
front and outer radio lobe already lie outside the cooling flow
region, ``wasting'' energy in the cluster at large radii and
making the inefficiency manifest.  Energy that is deposited outside
the cooling flow region adds to the total energy of the ICM and, as
noted by \citet{dnm01}, it can make an appreciable contribution to
cluster ``preheating.''  For a cluster temperature of 4 keV, the
virial mass of Hydra A is about $5.4\times10^{14}\ \msun$ \citep{bn98}
and its gas mass about $ 7.6\times10^{13}\ \msun$ (for a gas fraction
of 14\%).  The mean heating power required to add 1 keV per particle
to this gas over $10^{10}$ y is $8\times10^{44}\ \ergps$.  Our
estimates of the mean heating power from AGN outbursts lie in the
range $6\times10^{44}$ -- $2\times10^{45}\ \ergps$, so that the
``waste'' power could easily be a major source of preheating in Hydra
A.  There is no reason to believe that outbursts in much smaller
systems \citep[\eg][]{fj01} are more efficient, so these may also make
significant contributions to preheating.

\citet{brm04} found that bubble heating alone is insufficient to make
up for radiative losses in most cooling flows.  It should be noted
that Hydra A was one of the exceptions.  The total of $pV$ for the two
1.4 GHz inner cavities in Hydra A is $1.7\times10^{59}\rm\ erg$
\citep{brm04}.  Taking the pressure from \citet{dnm01}, for cavity C
to the northeast we estimate $pV \simeq 3.7\times10^{59}\rm\ erg$.
Doubling this to allow for outer cavities to the southwest, the total
of $pV$ for cavities in Hydra A is now $\sim 9\times10^{59}\rm\ erg$.
If the cavities are filled with relativistic plasma, their total
enthalpy is the sum of $4 pV$ for each, roughly one third of the
estimated total energy of the shock.  The shocks in Hydra A and other
systems \citep{fsa03,fnh04} show that bubble enthalpy may not be the
most significant form of mechanical energy produced by AGN outbursts.
As noted above, two systems have been found with even more energetic
shocks \citep[Hercules A, Nulsen \etal, in preparation; MS0735.6+7421,
][]{mnw04}.  If they are a general feature of AGN outbursts, shocks
will certainly help to account for the energetics of cooling flows.
\citet{brm04} found a large scatter in the ratio of radio power to
bubble power, suggesting that current radio power is not a good
measure of the history of AGN outbursts, even on timescales $\sim10^7$
-- $10^8$ y.  However, while the central galaxies in cooling flow
clusters are frequently active in the radio, radio sources as large as
Hydra A are exceptional.  If such outbursts do occur in a significant
proportion of all clusters, they can only be powerful radio sources
for a relatively small fraction of the time.

\subsection{Black Hole Growth}

The mass, $M_{\rm s}$, that was accreted by the AGN to fuel this
outburst can be related to the shock energy by $E_{\rm s} = \epsilon
M_{\rm s} c^2$, where $\epsilon$ is the efficiency of the AGN for
converting accreted mass into jet power (or other mechanical output).
With our value for the shock energy, this gives $M_{\rm s} \simeq 5
\times 10^6 \epsilon^{-1}\ \msun$.  This exceeds the estimated
mass of the black hole in Hydra A, $M_{\rm h} \simeq 4\times 10^9\
\msun$ \citep{sce00}, unless the mechanical efficiency of the AGN
$\epsilon \gtrsim 10^{-3}$.  The mean mechanical power of the
outburst, $P_{\rm s} \simeq 2\times10^{45}\ \ergps$, is in the
range of quasar luminosities, but a modest fraction of the Eddington
limit for such a massive black hole.  It is well in excess of the
power needed to make up for radiative losses from the cooling flow.

\section{Conclusion} \label{sec:conc}

A deep \chandra\ X-ray image of Hydra A has revealed new structure
associated with the Hydra A radio source.  This includes a break in
the surface brightness 200 -- 300 kpc from the cluster center that we
interpret as a shock front.  The shock front surrounds the low
frequency radio lobes, closely in places, showing a significant
resemblance, and we interpret it as the cocoon shock of the radio
source.

The Mach number of the shock varies in the range $\sim 1.2$ -- 1.4
around the front.  Simple modeling shows that the shock outburst
commenced $\sim 1.4\times10^8$ y ago and its energy
$\sim 9\times10^{60}$ erg.  The mean mechanical power of the outburst
$\sim 2\times10^{45}\ \ergps$, in the range of quasar luminosities.
Such large outbursts can easily account for the radiative losses from
the cooling flow in Hydra A.  If they are part of a feedback cycle
that sustains the cool gas in Hydra A, then the feedback heating
process is inefficient, in the sense that more than half of the
mechanical energy output of the AGN is deposited in regions outside
the cooling flow.  Over time this will have produced significant
``preheating'' of Hydra A.

\acknowledgments

PEJN was partly supported by NASA grants NAS8-01130 and NAS8-03060.
We thank Maxim Markevitch for drawing our attention to the the shock
feature before the new data were taken.  We gratefully acknowledge the
assistance of Alexey Vikhlinin and Maxim Markevitch in reducing the
\chandra\ data.  We thank Wendy Lane Peters and Namir Kassim for
providing the radio map.


\end{document}